\documentclass[epjCONF]{svjour}
\usepackage{amsmath}
\usepackage{graphicx}
\usepackage[varg]{txfonts} 
\usepackage[latin1]{inputenc}
\usepackage{bm}

\session-title{%
19$^{\textnormal{\footnotesize th}}$ International %
IUPAP Conference on Few-Body Problems in Physics%
}
\begin{document}
\title{%
Hyperfine Interaction in Quarkonia
}%
\author{%
Kamal K. Seth\fnmsep\thanks{\email{kseth@northwestern.edu}} 
}
\institute{%
Northwestern University, 2145 Sheridan Rd., Evanston, IL 60201, USA
}
\abstract{
The recent experimental developments in the measurement of hyperfine splittings in the bound states of charmonium and bottomonium are presented.  Their implications for the hyperfine interactions in the heavy quark systems are discussed.
} 
\maketitle
%
%
%
\section{Introduction}

The richness of the spectra of the excited states of atoms as well as hadrons lies not only in the principal quantum number and angular momentum dependence of the states, but in the \textbf{spin--dependent} multiplicities.  These arise from spin--orbit, tensor, and spin--spin interactions between the constituents.  Of these three, the most interesting is the \textbf{hyperfine} structure that arises due to the magnetic interactions between the spins, which causes the splitting between \textbf{spin--singlet} ($s_1+s_2=s=0$) and \textbf{spin--triplet} ($s_1+s_2=s=1$) states.  It is the transition between the $^3S_1$ and $^1S_0$ states of the hydrogen atom which gives rise to the famous \textbf{21~cm line} which is the workhorse of microwave astronomy.

Hyperfine interaction is equally important in hadron spectroscopy.  In the quark model, the ground state singlet masses of mesons made up of two quarks are simply given by
\begin{gather}
M(q_1\bar{q}_2)=m_1(q_1)+m_1(q_2)+\frac{32\pi\alpha_S}{9m_1m_2} |\psi(0)|^2 \vec{s}_1\cdot\vec{s}_2\\
\nonumber \mathrm{where}~\left<\vec{s}_1\cdot\vec{s}_2\right>=-\frac{3}{4}~~~\mathrm{for}~s=0,~~=+\frac{1}{4}~~~\mathrm{for}~s=1
\end{gather}
The hyperfine splitting is
\begin{equation}
\Delta M_{hf}(nS)\equiv M(n^3S_1) - M(n^1S_0) = \frac{32\pi\alpha_S}{9m_1m_2} |\psi_n(0)|^2
\end{equation}

It is remarkable how well this textbook prediction works with the rather realistic assumption about the strong coupling constant $\alpha_S(u,d)=0.6$, $\alpha_S(u,d,s)=0.4$, $\alpha_S(c)=0.32$, $\alpha_S(b)=0.18$, and that $|\psi(0)|^2/m_1m_2$ is a constant, $=33$.  As shown in Table~I, the predicted hyperfine splittings, $\Delta M(^3S_1-^1S_0)$ agree very well with their experimental values.

\begin{table*}[tb]
\caption{Hyperfine splittings $\Delta M_{hf}=M(^3S)-M(^1S)$ for $q\bar{q}$ mesons.}
\begin{center}
\begin{tabular}{lccccc}
\hline
$\Delta M_{hf}(1S)$ & ($\eta,\omega$) & ($D,D^*$) & ($D_s,D_s^*$) & ($\eta_c,J/\psi$) & $(\eta_b,\Upsilon(1S))$ \\
\hline
Eq.~1 (MeV) &  $221$ &  $147$ &  $147$ &  $118$ &  $66$ \\
Expt. (MeV) &  $234$ &  $142$ &  $144\pm1$ &  $117\pm1$ &  $71\pm4$ \\
\hline
\end{tabular}
\end{center}
\begin{small}
$^*$ With the same parameters the ($\pi,\rho$) and ($K,K^*$) splittings are both predicted to be a factor three smaller than observed.  This may be attributed to the small masses of the Goldstone bosons  $\pi$ and $K$.
\end{small}
\end{table*}

\section{Hyperfine Interaction Between Heavy Quarks}

The hyperfine splittings of heavy quarkonia require a more careful study than their inclusion in Table~I might suggest.

The analogy we have made between atoms and mesons, which implies a Coulombic or ($\propto1/r$) interaction, is not quite correct.  The $q\bar{q}$ central potential is known to have an \textbf{additional confinement part} which is not well understood, even though it is generally assumed to be a Lorentz scalar, and is parameterized as being proportional to $r$
\begin{equation}
 V(q\bar{q})=\frac{4}{3}\frac{\alpha_S}{r} + Cr
\end{equation}
This is the famous Cornell potential, illustrated in Fig.~1.

\subsection{Effect of Quark Confinement}

The Coulombic, $1/r$ vector part of the Cornell central potential implies the familiar spin--dependent interactions, the spin--orbit, the tensor, and the spin--spin interactions.  The contribution of the confinement potential to spin--dependence does not follow easily.  If it is \textbf{assumed} to be Lorentz scalar as is normally done, it can not contribute to the hyperfine interaction.  However, it may very well have a different Lorentz character.  Only experiments can decide.  Further, the presence of a confinement interaction poses the additional question of how the hyperfine interaction changes with the radius of the potential at which the specific states reside.  This is illustrated in Fig.~1 by vertical lines marking the approximate radii corresponding to the $\left|c\bar{c}\right>$ and $\left|b\bar{b}\right>$ bound states.  Clearly, the importance of the confining potential changes.  The interesting questions therefore are how  the consequent hyperfine splitting, changes with the
\begin{enumerate}
\item principle quantum number, i.e., between 1S and 2S states
\item angular momentum, i.e. from $L=0$ (S--states) to $L\ne0$ (e.g., P--states)
\item quark masses, e.g., from $c$--quark to $b$--quark states,
\end{enumerate}
and what the changes imply for the spin--spin hyperfine interaction between heavy quarks.
The best place to address these questions is with hidden flavor ($q$ and $\bar{q}$ of the same flavor) mesons, such as the \textbf{charmonium} and \textbf{bottomonium} mesons.

\begin{figure}
\includegraphics[width=\columnwidth]{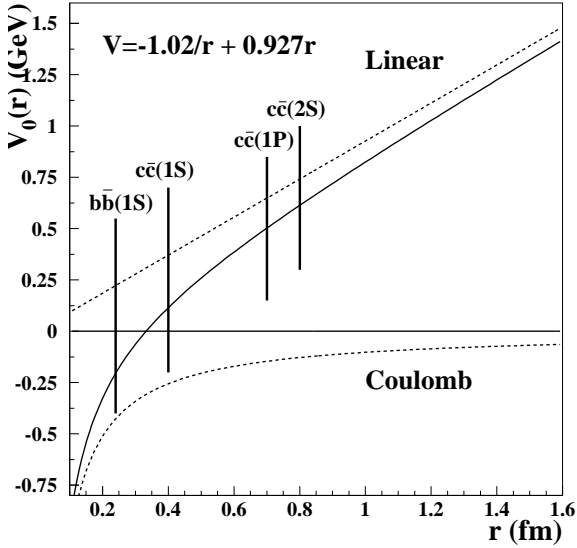}
\caption{The phenomenological $q\bar{q}$ Cornell potential.  The vertical lines show the approximate location of the $\left|c\bar{c}\right>$ charmonium and $\left|b\bar{b}\right>$ bottomonium bound states.}
\end{figure}

\subsection{The Experimental Problem}

There is a generic problem in measuring hyperfine splittings, 
$$\Delta M_{hf}(nL)\equiv M(n^3L) - M(n^1L).$$  
The problem is that while the triplet states are conveniently excited in $e^+e^-$ annihilation, either directly (e.g., $^3S_1$) or via \textbf{strong E1} radiative transitions (e.g., $^3P_J$), the excitation of singlet states is either forbidden, or possible only with \textbf{weak M1} allowed ($n\to n$) and forbidden ($n\to n'$) transitions.  This results in the following situation.

While the $J/\psi(1^3S_1)$ state of charmonium was discovered in 1974~\cite{jpsi-disc} and $\Upsilon(1^3S_1)$ state of bottomonium was discovered in 1977~\cite{ups-disc}.  After several false identifications, $\eta_c(1^1S_0)$ was identified in 1980~\cite{etac-disc}, and for more than thirty years, the only hyperfine splitting which had been measured in a hidden flavor meson was~\cite{pdg}
\begin{equation}
\Delta M_{hf}(1S)_{c\bar{c}}\equiv M(J/\psi)-M(\eta_c) = 116.6\pm1.0~\mathrm{MeV}
\end{equation}
No other singlet states, 
$$\eta_c'(2^1S_0)_{c\bar{c}},~~~h_c(1^1P_1)_{c\bar{c}},~~\mathrm{or}~~\eta_b(1^1S_0)_{b\bar{b}}.$$ 
were identified, and none of the important questions about the hyperfine interaction which we posed earlier could be addressed.

This has changed in the last few years.

I want to briefly describe these recent experimental developments and their consequence for theory.

\subsection{Hyperfine Splitting in Charmonium Radial Excitation: The Search for $\eta_c'(2^1S_0)$}

For charmonium, the hyperfine splitting in this case is 
\begin{equation}
\Delta M_{hf}(2S)_{c\bar{c}}\equiv M(\psi'(2^3S_1))-M(\eta_c'(2^1S_0))
\end{equation}
The mass of the triplet state radial, $\psi'(2^3S_1)$, is very well measured, $M(\psi'(2^3S_1))=3686.09\pm0.04$~MeV~\cite{pdg}.  What is required is to identify $\eta_c'(2^1S_0)$, and to measure its mass with precision.  The identification of $\eta_c'$ in the radiative decay
\begin{equation}
\psi'(2^3S_1)\to\gamma\eta_c'(2^1S_0)
\end{equation}
is difficult because the transition is a weak M1, and is predicted to have very low energy ($E_\gamma\sim30-50$~MeV).  It has never been identified!  One has to find other ways of populating $\eta_c'$, for example in photon--photon fusion, or in $B$--decays, and to reconstruct it in some of its hadronic decays.  So far only one such decay has been identified.  It is
\begin{equation}
\eta_c'(2S)\to K_SK\pi.
\end{equation}

\begin{figure}
\includegraphics[width=3.in]{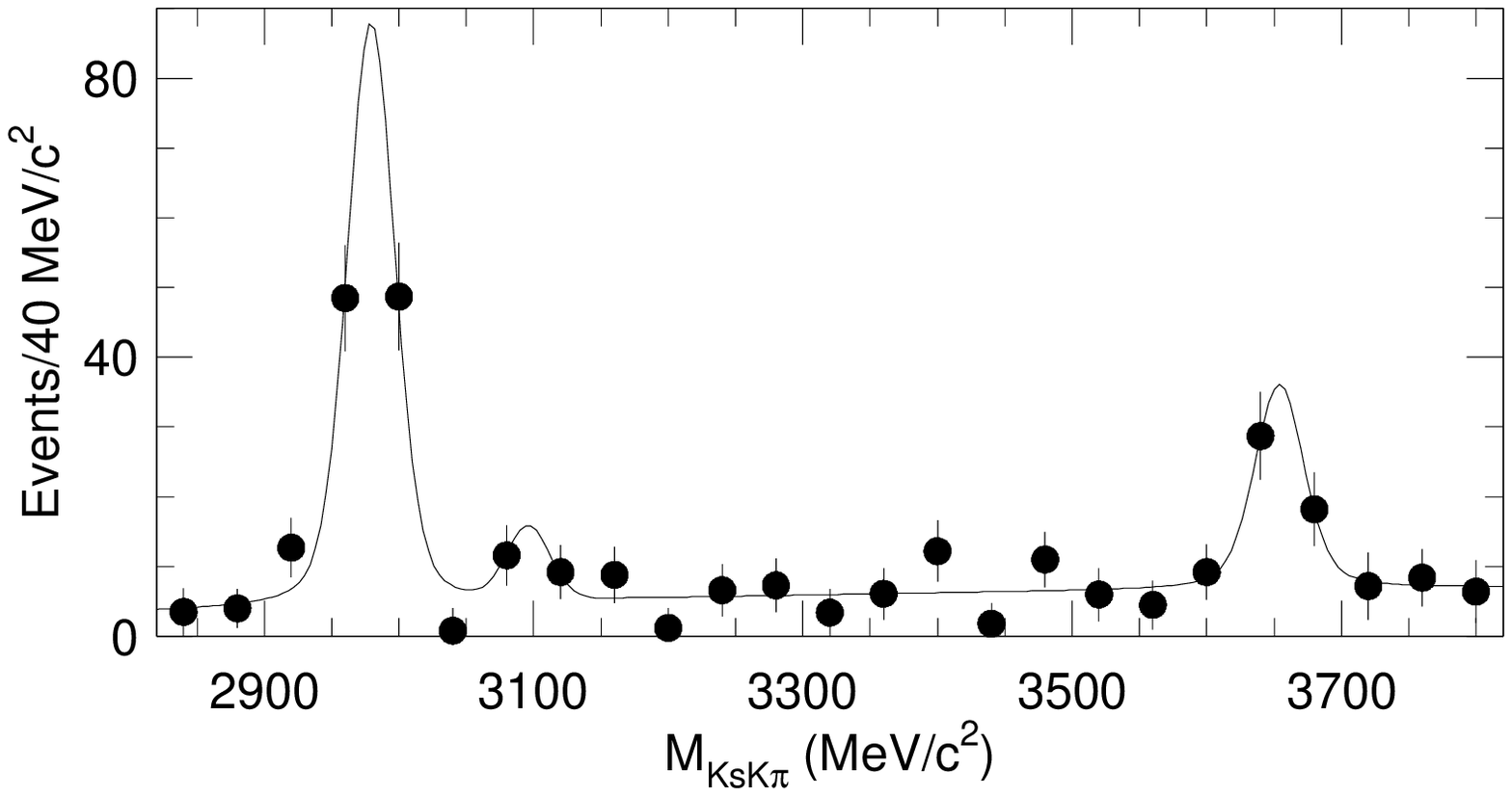}

\hspace*{-10pt}\includegraphics[width=3.25in]{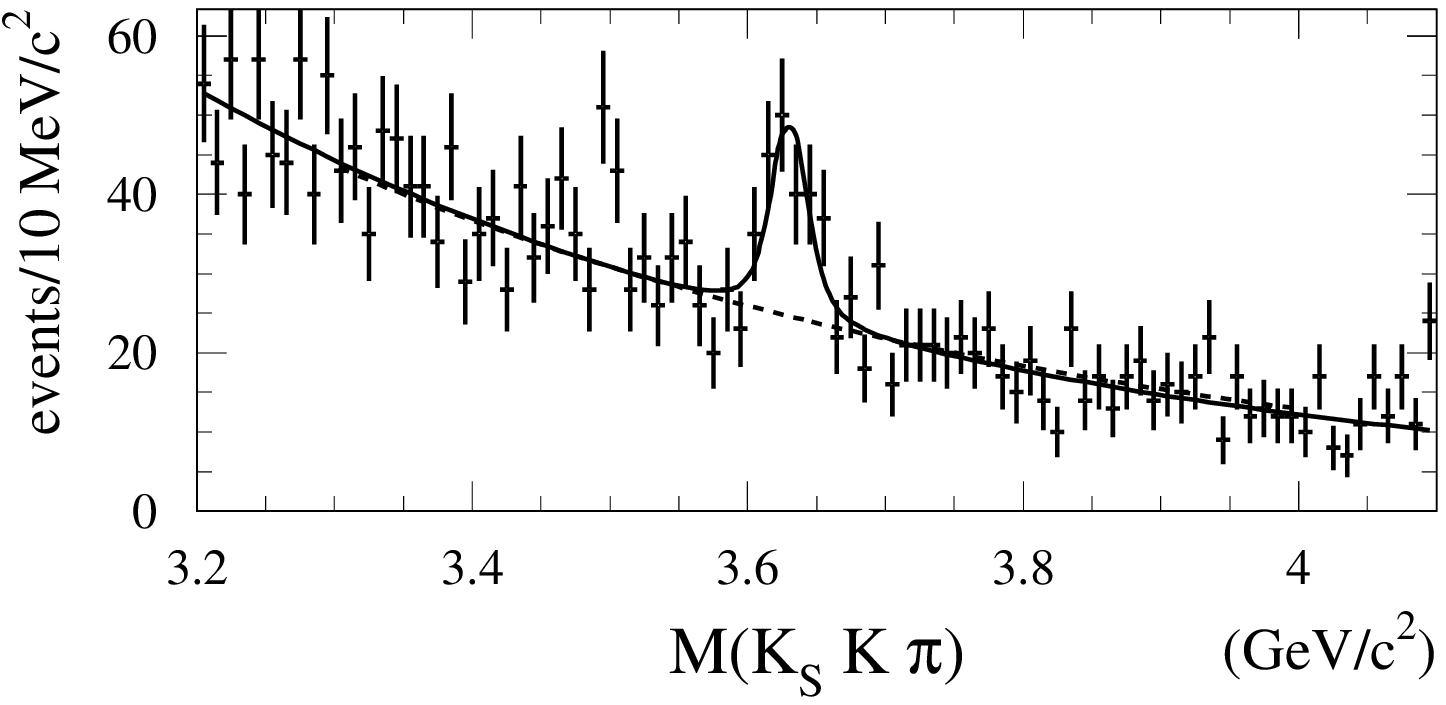}

\vspace*{-35pt}

\hspace*{-5pt}\includegraphics[width=3.5in]{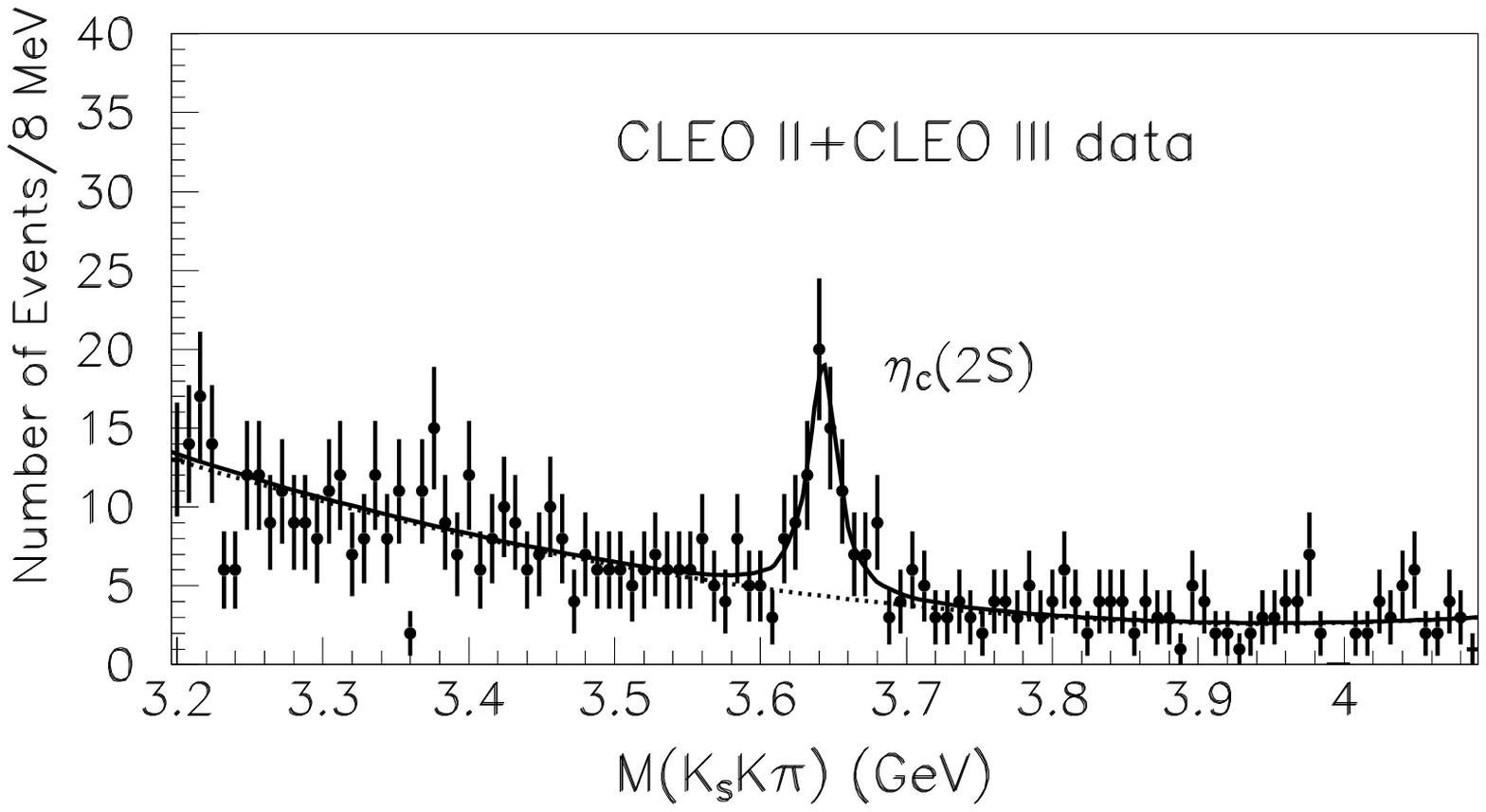}
\caption{Identification of $\eta_c'(2^1S_0)$ in $K_SK\pi$ invariant mass, (top) from $B$ decays by Belle, (middle) from two~photon fusion by BaBar, and (bottom) from two~photon fusion by CLEO.}
\end{figure}

The first identification of $\eta_c'$ came from an unexpected source.  In 2002, Belle reported $\eta_c'$ observation in two different measurements using 42~fb$^{-1}$ of $e^+e^-$ annihilation at $\sqrt{s}=10.58$~GeV.  In $B$ decays into $K(K_SK\pi)$ they claimed its identification with 56 counts in $K_SK\pi$, and $M(\eta_c')=3654\pm10$~MeV~\cite{belle-etacpb}.  In double charmonium production, $J/\psi\cdot\eta_c'$ they observed a signal with $42^{+15}_{-13}$ counts, and $M(\eta_c')=3622\pm12$~MeV~\cite{belle-etacpcc}.  The need to confirm the discovery of $\eta_c'$, and to resolve the discrepancy of 32~MeV between the two masses motivated us at CLEO to search for $\eta_c'$ in two--photon fusion with 27~fb$^{-1}$ of data taken by us in the $\Upsilon$ region.  We reported observation in $K_SK\pi$ decay with $61^{+19}_{-15}$ events and $M(\eta_c')=3642.9\pm3.4$~MeV, and consequently $\Delta M_{hf}(2S)_{c\bar{c}}=43.1\pm3.4$~\cite{cleo-etacp}.  

Our observation was followed by a similar measurement by BaBar with 88~fb$^{-1}$ of data taken at $\sqrt{s}=10.58$~GeV.  They observed $112\pm24$~events, and reported $M(\eta_c')=3630.8\pm3.5$~MeV~\cite{babar-etacp}.  The identification of $\eta_c'$ was firmly established, albeit with uncomfortably large differences in mass.  The width of $\eta_c'$ remains uncertain within a factor~2 even to this day.  The three measurements are illustrated in Fig.~2.  Since then both Belle and BaBar have reported more mass measurements, and the PDG08~\cite{pdg} average of all the mass measurements is $M(\eta_c')=3537\pm4$~MeV, which leads to the hyperfine splitting
\begin{equation}
\Delta M_{hf}(2S)_{c\bar{c}} = 49\pm4~\mathrm{MeV}
\end{equation}
Recall that $\Delta M_{hf}(1S)_{c\bar{c}} = 116.6\pm1.0~\mathrm{MeV}$~\cite{pdg}.

There~~ are~~ numerous~~ pQCD--based~~ predictions~~ for $\Delta M_{hf}(2S)_{c\bar{c}}$, and they range all over the map (and occasionally even hit 50~MeV).  However, it is fair to say that nobody expected the 2S hyperfine splitting to be $\sim2.5$ times smaller than the 1S hyperfine splitting.  A model--independent prediction, relating 2S to 1S splitting using $J/\psi(1S)$ and $\psi'(2S)$ masses, and $e^+e^-$ decay widths, gives $\Delta M_{hf}(2S)_{c\bar{c}}=68\pm7$~MeV, which is also off the mark.

So far lattice calculations are not of much help.  The two predictions based on unquenched lattice calculations are
\begin{align}
\mathrm{Columbia}~[9]: & \quad \Delta M_{hf}(2S)_{c\bar{c}}=75\pm44~\mathrm{MeV} \\
\mathrm{CP-PACS}~[10]: & \quad \Delta M_{hf}(2S)_{c\bar{c}}=25=43~\mathrm{MeV} & 
\end{align}
It has been suggested that the smaller than expected 2S hyperfine splitting is a consequence of $\psi(2S)$ being very close to the $\left|c\bar{c}\right>$ break-up threshold, and continuum mixing lowers its mass, resulting in a reduced difference $M(\psi'(2S))-M(\eta_c'(2S))$.  However, no definitive numerical predictions are available so far.

\subsection{Hyperfine Splitting in Charmonium P--wave: The Search for $h_c(^1P_1)$}

In this case, we have a very simple, and provocative theoretical expectation, namely
\begin{equation}
\Delta M_{hf}(1P) \equiv M(^3P) - M(^1P) = 0
\end{equation}
This arises from the fact that a non-relativistic reduction of the Bethe-Salpeter equation makes the hyperfine interaction a \textbf{contact interaction}.  Since only S--wave states have finite wave function at the origin,
\begin{equation}
\Delta M_{hf}(L\ne0)=0.
\end{equation}
We can test this prediction in charmonium by
\begin{itemize}
\item identifying the singlet--P state $h_c(1^1P_1)$, and
\item by estimating $M(^3P)$, given the masses of the triplet--P states $\chi_{0,1,2}~(^3P_{0,1,2})$
\end{itemize}

\begin{figure}
\includegraphics[width=\columnwidth]{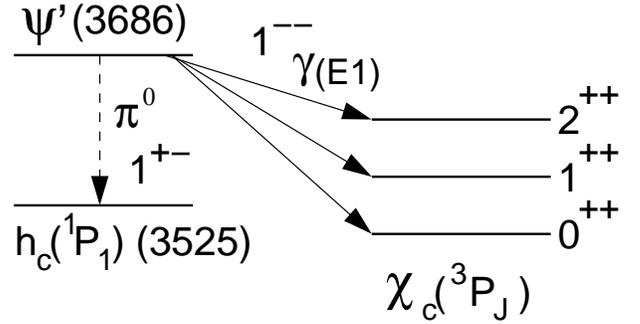}
\caption{Comparing allowed E1 transitions from $\psi'(^3S_1)$ to $\chi_{cJ}(^3P_J)$ states of charmonium with the isospin forbidden $\pi^0$ transition to the singlet P--state $h_c(^1P_1)$.}
\end{figure}

The experimental identification of $h_c(1^1P_1)$ is even more difficult than that of $\eta_c'$.  The centroid of the $^3P_J$ states is at $3525.30\pm0.04$~MeV~\cite{pdg}. If Eq.~11 is true, $M(h_c)\approx3525$~MeV, i.e., $\sim160$~MeV below the $\psi'(2S)$ state from which it must be fed.  Unfortunately, populating $h_c$ has problems.
\begin{itemize}
\item The radiative transition $\psi'(1^{--})\to\gamma h_c(1^{+-})$ is forbidden by \textbf{charge conjugation} invariance.  
\item The only other alternative is to populate $h_c$ in the reaction $\psi'\to\pi^0h_c$.
But that is not easy, because a $\pi^0$ transition ($M(\pi^0)=139$~MeV) has very little phase space, and further, the reaction is forbidden by \textbf{strict isospin conservation}. Nevertheless, this is the only possible way of populating $h_c$, and we at CLEO had to valiantly go for it.
\end{itemize}

As with all difficult searches, there is a history of $h_c(^1P_1)$ searches.  In 1982, the Crystal Ball Collaboration searched for $h_c$ in inclusive $\psi'\to\pi^0h_c$ with 0.9 million $\psi'$.  They found no evidence of $h_c$ in the mass region $M(h_c)=3440-3535$~MeV, and established the upper limit  $\mathcal{B}(\psi'\to\pi^0h_c)<1.09\%$~(95\%~CL)~\cite{cb-hc}.

In 1992, the Fermilab E760 Collaboration, taking advantage of the fact that, in contrast to $e^+e^-$ annihilation, $h_c$ can be directly formed in $p\bar{p}$ annihilation, searched for $h_c$ in $p\bar{p}\to h_c \to \pi^0J/\psi$.  This time there was plenty of phase space for $\pi^0$, but it was the decay which was isospin violating.  E760 scanned the region $\sqrt{s(p\bar{p})}=3522.6-3527.2$~MeV with an integrated luminosity $\mathcal{L}(p\bar{p})=16$~pb$^{-1}$.
They reported~\cite{e760-hc} a statistically significant enhancement with $\sim30$ counts, which they attributed to $h_c$, with 
\begin{equation}
M(h_c)=3526.2\pm0.15\pm0.20~\mathrm{MeV}.
\end{equation}

In 2005, the Fermilab E835 Collaboration, repeated their search for $h_c$ with three times larger luminosity ($\mathcal{L}=48$~pb$^{-1}$) by combining their data for 1997 and 2000 runs.  In the reaction $p\bar{p}\to h_c \to \pi^0J/\psi$, no evidence for $h_c$ the region in $M=3520-3540$~MeV was found, in contrast to their 1992 report.  However, they reported~\cite{e835-hc} that in the reaction, $p\bar{p}\to h_c \to \gamma\eta_c$, an enhancement consisting of 13 counts with significance $\sim3\sigma$, and mass
\begin{align}
M(h_c)=3525.8\pm0.2\pm0.2~\mathrm{MeV}, 
\end{align}

\begin{figure}
\includegraphics[width=\columnwidth]{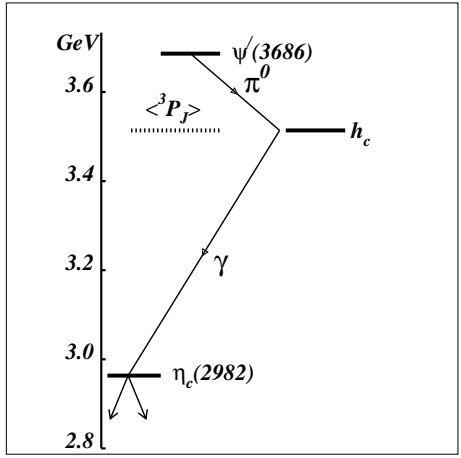}
\caption{Illustrating the sequence of decays $\psi'(2^3S_1)\to\pi^0h_c(1^1P_1)$, $h_c\to\gamma\eta_c(1^1S_0)$ used by CLEO to identify $h_c$.}
\end{figure}

In 2005, we at CLEO made the first firm identification (significance$>6\sigma$) of $h_c$ in the reaction 
$$\psi'\to\pi^0h_c,~~h_c\to\gamma\eta_c$$ 
which is illustrated in Fig.~4.  In an analysis of 3.08~million $\psi'$ decays $h_c$ was identified~\cite{cleo-hc} with $N(h_c)=178\pm40$~events,
\begin{gather}
M(h_c)=3524.4\pm0.6\pm0.4~\mathrm{MeV},\\
\nonumber\mathcal{B}_1(\psi'\to\pi^0h_c)\times\mathcal{B}_2(h_c\to\gamma\eta_c)=(4.0\pm0.8\pm0.7)\times10^{-4}.
\end{gather}  

\begin{figure}
\includegraphics[width=\columnwidth]{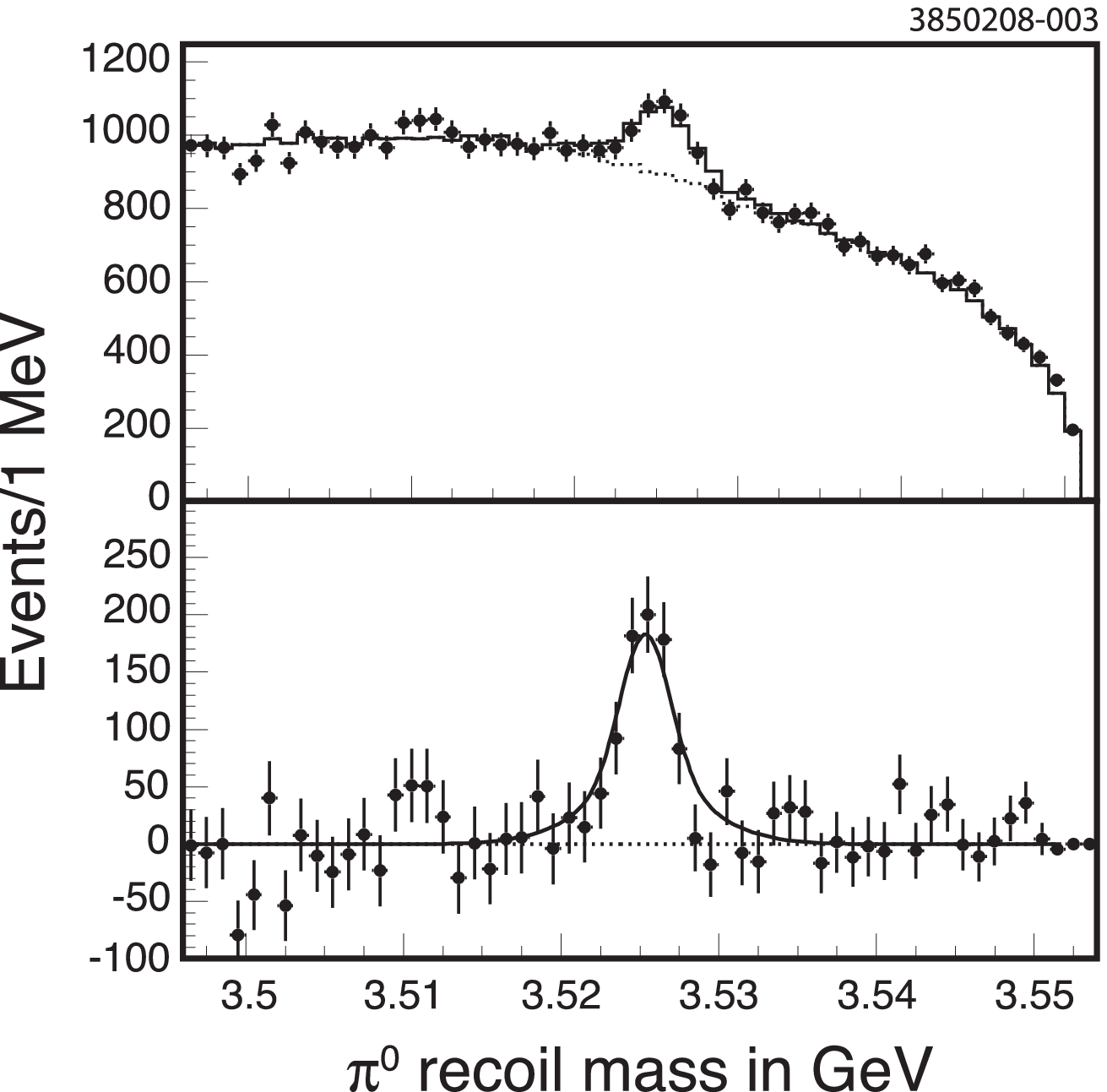}

\includegraphics[width=\columnwidth]{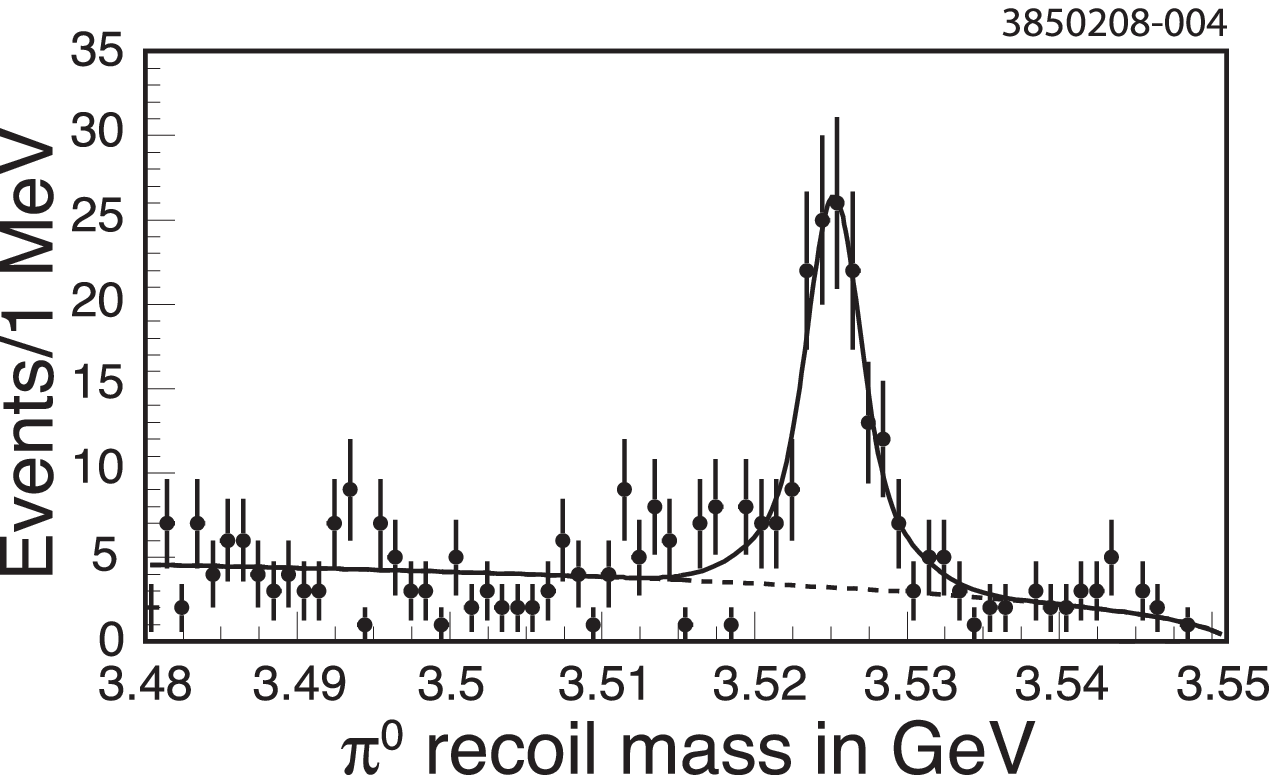}
\caption{The recoil mass of $\pi^0$ in the decay $\psi'\to\pi^0h_c$.  (Top) Full and background subtracted spectra for inclusive analysis.  (Bottom) Spectrum of exclusive analysis.}
\end{figure}

In 2008, we repeated our measurement with 8 times larger luminosity, and  24.5 million $\psi'$~\cite{cleo-hc-new}.  As before, data were analyzed in two ways.  In the inclusive analysis, the photon energy, $E_\gamma$, was loosely constrained, but the decay products of $\eta_c$ were not identified. In the exclusive analysis, instead of constraining $E_\gamma$ fifteen hadronic decay channels of $\eta_c$ were measured. 
As shown in Fig.~5, a total of $N(h_c)=1282\pm119$~events ($1146\pm118$ from inclusive analysis, and $136\pm14$ from exclusive analysis) were observed with significance~~$>13\sigma$. Precision results were obtained
\begin{align}
M(h_c) & =3525.28\pm0.19\pm0.12~\mathrm{MeV},\\
\nonumber\mathcal{B}_1\times\mathcal{B}_2 & =(4.19\pm0.32\pm0.45)\times10^{-4}.
\end{align}
 $h_c(^1P_1)$ \textbf{is now firmly established.}

If it is assumed that $M(^3P)$ is identical to the centroid of the triplet--P states, $\left<M(^3P_J)\right>=[5M(\chi_{c2})+3M(\chi_{c1})+M(\chi_{c0})]/9=3525.30\pm0.04$~MeV, then the above $M(h_c)$ leads to the hyperfine splitting,
\begin{equation}
\Delta M_{hf}(1P)_{c\bar{c}}=\left<M(^3P_J)\right>-M(^1P_1)=0.08\pm0.22~\mathrm{MeV}
\end{equation}
But, $\left<M(^3P_J)\right>_{0,1,2} \ne M(^3P)$!

The centroid  $\left<M(^3P_J)\right>$ is a good measure of $M(^3P)$ only if the spin--orbit splitting between the states $^3P_2$, $^3P_1$, and $^3P_0$ is perturbatively small.  It is obviously not so.
The splitting, $M(^3P_2)-M(^3P_0)=141.5$~MeV, is hardly small.
Further, the perturbative prediction is that 
\begin{align}
M(^3P_1)-M(^3P_0)& =\frac{5}{2}\left[M(^3P_2)-M(^3P_1)\right]\\
\nonumber   & =113.9\pm0.3~\mathrm{MeV},
\end{align}
while the experimental value is
\begin{equation}
M(^3P_1)-M(^3P_0)=95.9\pm0.4~\mathrm{MeV}
\end{equation}
This is a 20\% difference! So we are obviously not in the perturbative regime.

This leads to serious questions.
\begin{itemize}
\item What mysterious cancellations are responsible for the wrong estimate of $M(^3P)$ giving the expected answer that
$$\Delta M_{hf}(1P)=0$$
\item Or, is it possible that the expectation is wrong?  Is it possible that the hyperfine interaction is not entirely a \textbf{contact interaction}?  
\item Potential model calculations are not of much help because they smear the potential at the origin in order to be able to do a Schr\"odinger equation calculation.
\item Can Lattice help?
\end{itemize}

\subsection{Lattice to the Rescue}

The Coulombic part of the $q\bar{q}$ interaction is vector, and in a non-relativistic reduction of the Bethe-Salpeter equation it leads to a contact spin--spin interaction, i.e., it predicts no long--range spin--spin interaction.  A long--range spin--spin interaction can be obtained either by considering a Lorentz vector part in the confinement potential, or by considering an extension of the one--gluon exchange for the central potential.

In a lattice calculation (unfortunately, still quenched), Koma and Koma~\cite{komakoma} address the latter possibility.  They claim clear evidence for deviations from a one--gluon vector exchange in the Bethe-Salpeter kernel which leads to a $\delta$--function spin--spin interaction.  They speculate on the introduction of a pseudo--scalar exchange (a $0^{-+}$ glueball?) in addition to the one--gluon vector exchange. As unconventional as this suggestion is, it is extremely important to confirm this.  Our measurement of the P--wave hyperfine splitting can hopefully shed light on the subject.

\section{Hyperfine Interaction Between $b$--Quarks: The Search for $\eta_b(^1S_0)_{b\bar{b}}$}

The $b\bar{b}$ bottomonium system is, in principle, the best one to study the fundamental aspects of the hyperfine interaction between quarks.  This is because the $b$--quarks are the heaviest quarks which make hadrons.  As a consequence, the quarks in bottomonium are far less relativistic ($(v/c)^2_{b\bar{b}}\approx0.08$ compared to $(v/c)^2_{c\bar{c}}\approx0.23$), and also have smaller strong coupling constant ($\alpha^S_{b\bar{b}}=0.18$ compared to $\alpha^S_{c\bar{c}}\approx0.35$).  Further, as shown in Fig.~1, the bottomonium states lie in the $q\bar{q}$ potential region dominated by the Coulombic part, and are therefore least affected by the uncertainties of the confinement part of the potential. All these properties make perturbative QCD more valid in bottomonium, and provide better testing ground for Lattice calculations.

 Unfortunately, until last year we had no knowledge of the hyperfine interaction between $b$--quarks.  The spin--triplet $\Upsilon(1^3S_1)$ state of bottomonium was discovered in 1977~\cite{ups-disc}, but its partner, the spin--singlet $\eta_b(1^1S_0)$ ground state of bottomonium, was not identified for thirty years, mainly for the same reasons we have mentioned before---the difficulty in observing weak M1 radiative transitions.  There were many pQCD based theoretical predictions  which varied all over the map, with $\Delta M_{hf}(1S)_b=35-100$~MeV, and $\mathcal{B}(\Upsilon(3S)\to\gamma\eta_b)=(0.05-25)\times10^{-4}$.  

So, we knew nothing about the hyperfine interaction between $b$--quarks.

This has changed now.  The $\eta_b(1^1S_0)$ ground state of the $\left|b\bar{b}\right>$ Upsilon family \textbf{has finally been identified!}

In July 2008, BaBar announced the identification of $\eta_b$~\cite{ups-babar}.  They analyzed the inclusive photon spectrum of
\begin{equation}
\Upsilon(3S)\to\gamma\eta_b(1S)
\end{equation}
in their data for \textbf{109 million} $\Upsilon(3S)$ (28~fb$^{-1}~e^+e^-$).  BaBar's success owed to their very large data set and a clever way of reducing the continuum background, a cut on the so--called thrust angle, the angle between the signal photon and the thrust vector of the rest of the event.  BaBar's results were:
\begin{gather}
M(\eta_b)=9388.9^{+3.1}_{-2.3}\pm2.7~\mathrm{MeV},\\
\nonumber \Delta M_{hf}(1S)_b=71.4^{+3.1}_{-2.3}\pm2.7~\mathrm{MeV} \\
\nonumber \mathcal{B}(\Upsilon(3S)\to\gamma\eta_b) = (4.8\pm0.5\pm1.2)\times10^{-4}
\end{gather}

Any important discovery requires independent confirmation.  At CLEO we had data for only \textbf{5.9~million} $\Upsilon(3S)$, i.e., about 20~times less than BaBar.    But we have better photon energy resolution, and we have been able to improve on BaBar's analysis technique.  We make three improvements.  We make very detailed analysis of the large continuum background under the very weak resonance photon peaks.  We determine photon peak shapes by analyzing background from peaks in background--free radiative Bhabhas and in exclusive $\chi_{b1}$ decays.  And we make a joint fit of the full data in three bins of $|\cos\theta_T|$, covering the full range $|\cos\theta_T|=0-1.0$.  So, despite our poorer statistics, we have succeeded in confirming BaBar's discovery.  
The results have since then been submitted for publication~\cite{ups-cleo-new}. The results agree with those of BaBar.

\begin{figure*}
\raisebox{3.37in}{\parbox{\columnwidth}{
\includegraphics[width=\columnwidth]{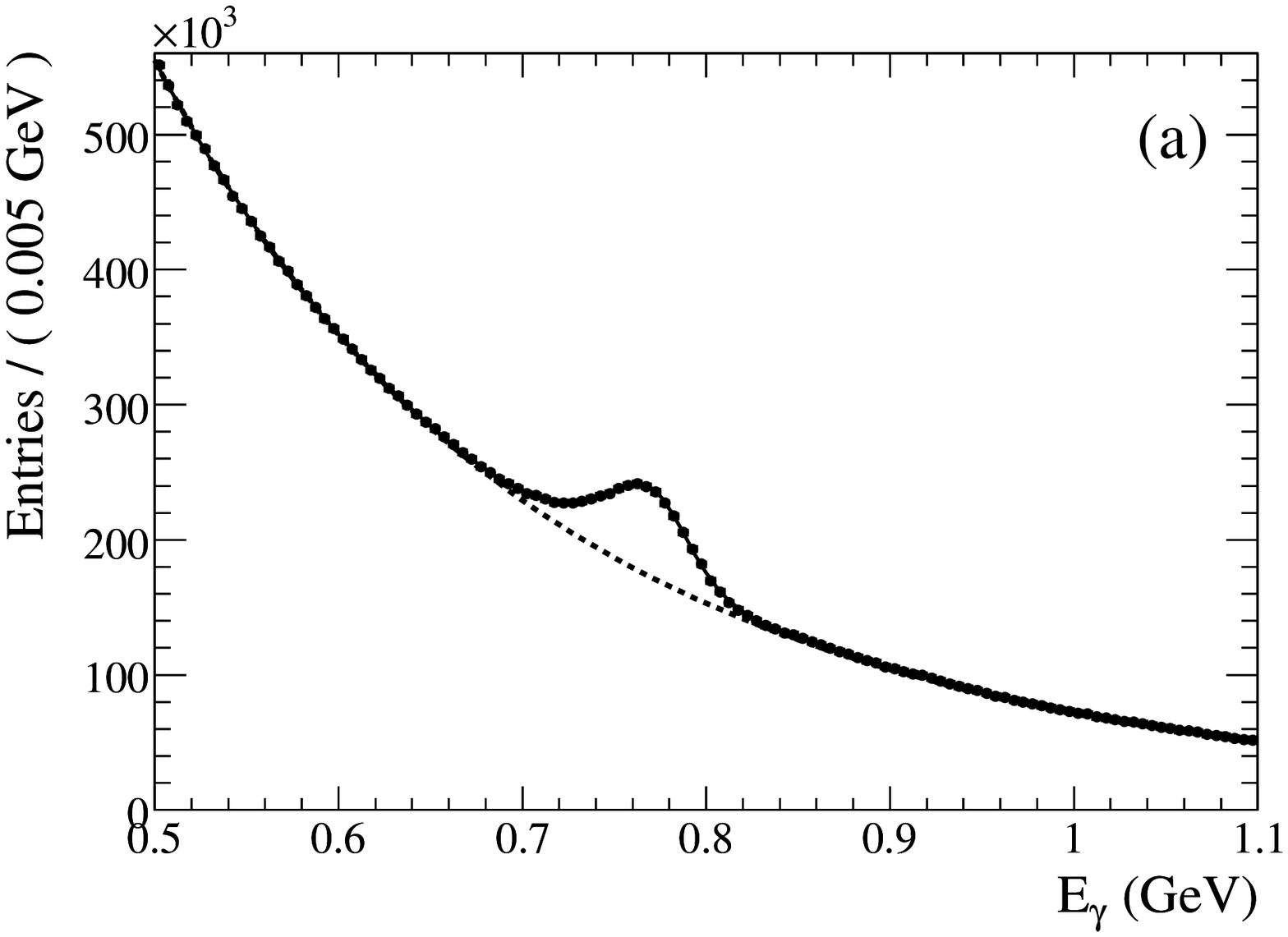}\\[4pt]
\includegraphics[width=\columnwidth]{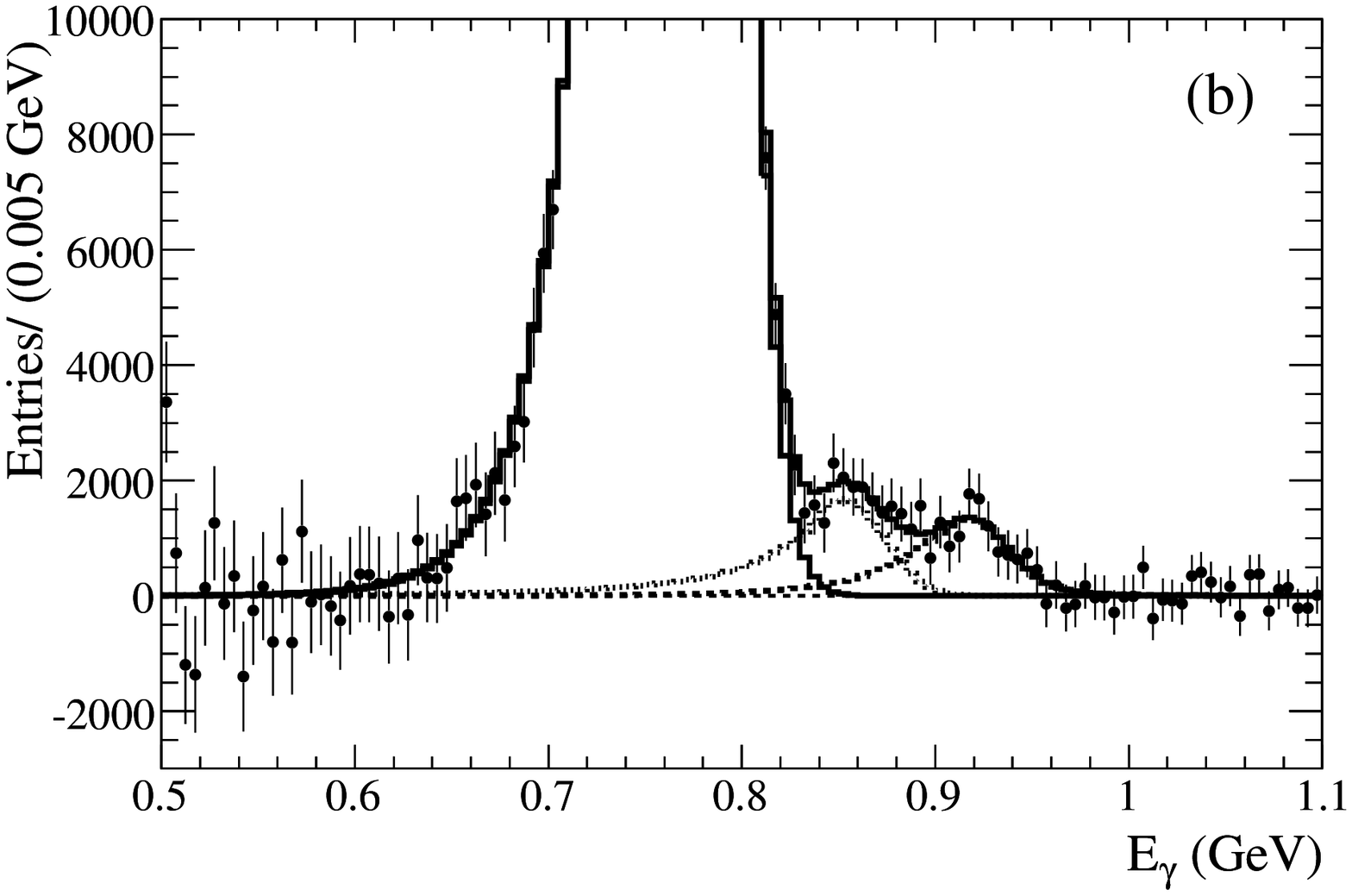}\\[4pt]
\includegraphics[width=\columnwidth]{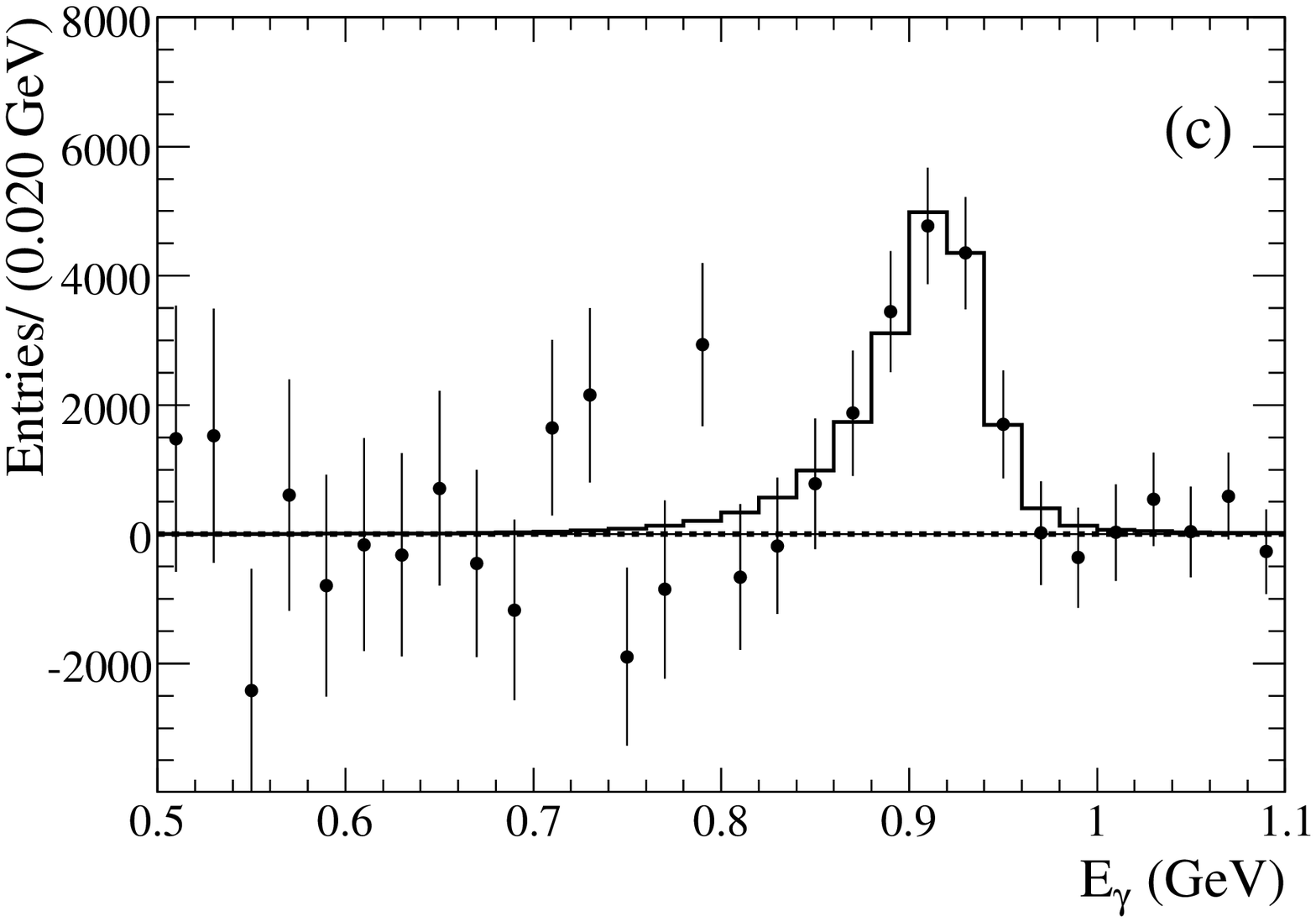}
}}
\includegraphics[width=\columnwidth]{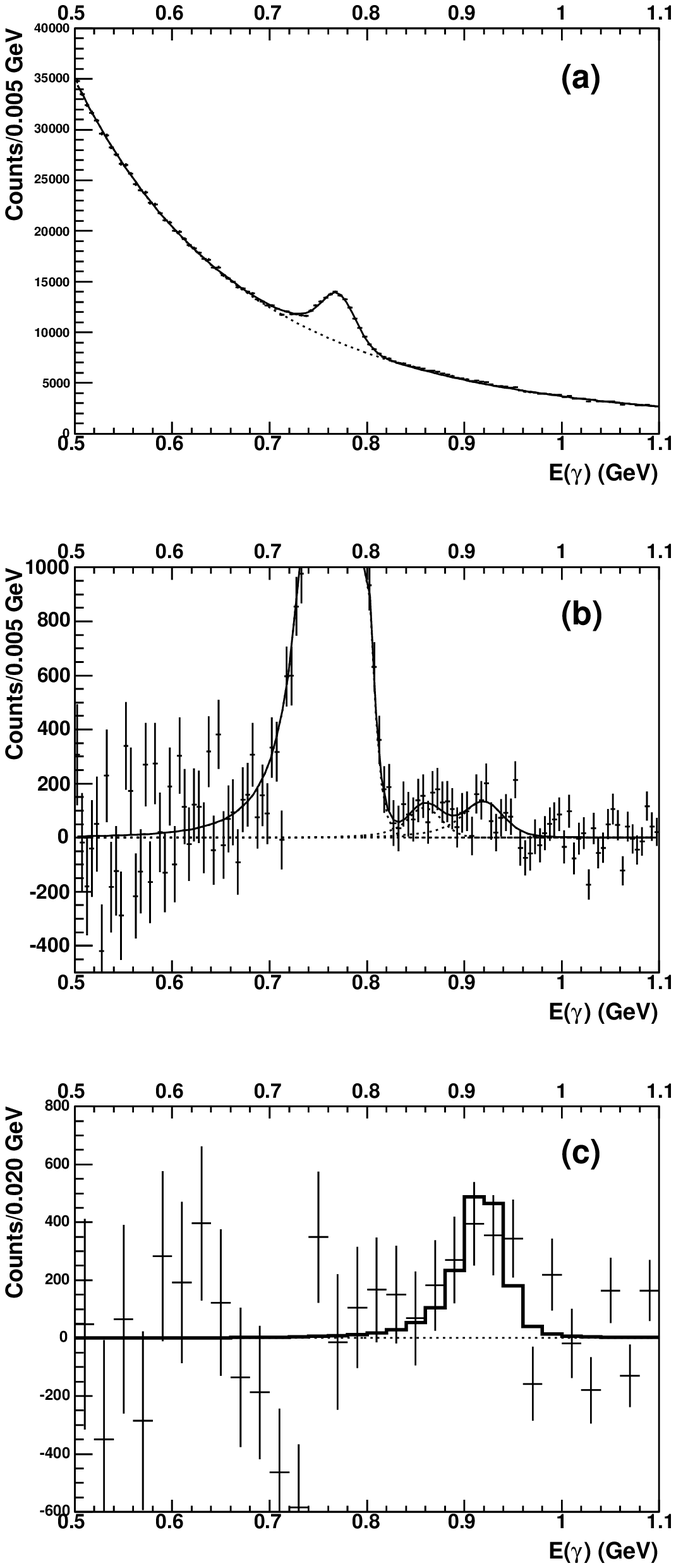}
\caption{Identification of $\eta_b$ in $\Upsilon(3S)\to\gamma\eta_b$ (left) from BaBar~\cite{ups-babar}, (right) from CLEO~\cite{ups-cleo-new}.  Panels (a) show the complete inclusive spectrum.  Panels (b) show the spectra after subtraction of the continuum background.  Panels (c) show after the $\chi_{bJ}$ and ISR peaks have also been subtracted.}
\end{figure*}

\begin{figure}
\includegraphics[width=\columnwidth]{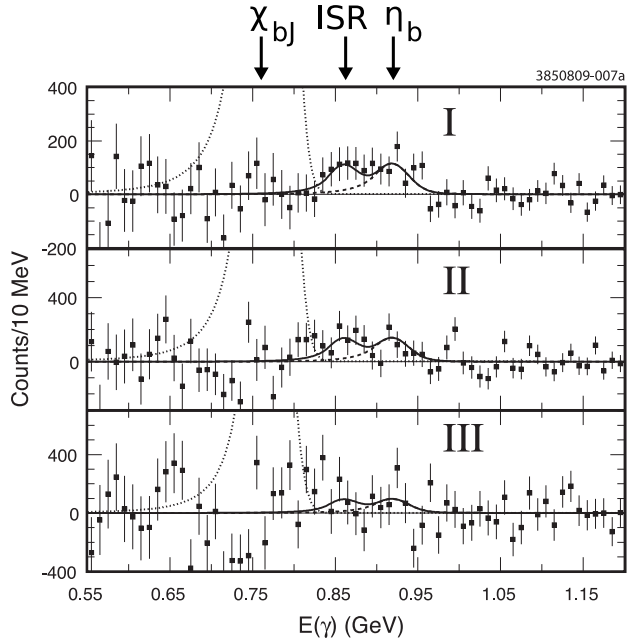}
\caption{Illustrating CLEO results for the identification of $\eta_b$ in a joint fit of data in three bins of the thrust angle,  I:~$|\cos\theta_T|=0-0.3$,  II:~$|\cos\theta_T|=0.3-0.7$,  III:~$|\cos\theta_T|=0.7-1.0$.}
\end{figure}

Our results are:
\begin{gather}
M(\eta_b)=9391.8\pm6.6\pm2.1~\mathrm{MeV}\\
\nonumber \Delta M_{hf}(1S)_b = 68.5\pm6.6\pm2.1~\mathrm{MeV}\\
\nonumber \mathcal{B}(\Upsilon(3S)\to\gamma\eta_b)=(7.1\pm1.8\pm1.2)\times10^{-4}
\end{gather}

The average of our and BaBar's result for the hyperfine splitting is
$$\left<\Delta M_{hf}(1S)_b\right> \equiv M(\Upsilon(1S))-M(\eta_b) = 70.6\pm3.5~\mathrm{MeV}$$
A recent unquenched lattice calculation predicts (NRQCD with $u,d,s$ sea quarks) $\Delta M_{hf}(1S)_b=61\pm14$~MeV.  A quenched lattice calculation (chiral symmetry and $s,c$ sea quarks) predicts $\Delta M_{hf}(1S)_b=70\pm5$~MeV.
Thus, as far as the hyperfine splitting for the $\left|b\bar{b}\right>$ is concerned, lattice calculations appear to be on the right track.

The situation is quite different with the predictions of the strength of the radiative transition, $\mathcal{B}(\Upsilon(3S)\to\gamma\eta_b(1S))$. There are no lattice predictions of transition strengths, so far, and pNRQCD predications for forbidden M1 transitions are orders of magnitude off~\cite{brambilla}.

\section{Summary}

To summarize, we now have well--measured experimental results for several hyperfine singlet/triplet splittings in heavy quark hadrons:\\[6pt]
\begin{tabular}{ll}
$\left|c\bar{c}\right>$ Charmonium: &   $\Delta M_{hf}(1S)=116.7\pm1.2$~MeV \\[2pt]
 &    $\Delta M_{hf}(2S)=43.2\pm3.4$~MeV \\[2pt]
 &    $\Delta M_{hf}(1P)=0.02\pm0.23$~MeV  \\[4pt]
$\left|b\bar{b}\right>$ Bottomonium: &   $\Delta M_{hf}(1S)=70.6\pm3.5$~MeV \\[6pt]
\end{tabular}

In charmonium, we do not have satisfactory understanding of the variation of hyperfine splitting for the S--wave radial states, and for P--wave states
\begin{itemize}
\item For charmonium, we do not have any unquenched lattice predictions, at present.
\item For bottomonium, lattice predictions are available, and they appear to be on the right track.
\item For neither charmonium or bottomonium there are any reliable predictions of transitions strength, particular for forbidden M1 transitions.
\end{itemize}

Much remains to be done.  On the experimental  front it is very important to identify for bottomonium the allowed M1 transition, $\Upsilon(1S)\to\gamma\eta_b(1S)$, and to identify the bottomonium singlet P--state, $h_b(^1P_1)$.  On the theoretical front one would like to see unquenched lattice calculations for charmonium singlets, and, of course, for transition strengths.

I wish to thank the U.S. Department of Energy for supporting the research reported here.

%

\end{document}